\documentclass[aps, prd, twocolumn, nofootinbib, longbibliography,showkeys, superscriptaddress]{revtex4-1}
\usepackage{cancel}
\usepackage{epsfig}
\usepackage[top=59pt,bottom=50pt,left=46pt,right=46pt]{geometry}
\usepackage{enumitem}
\usepackage{amsmath}
\usepackage{amssymb}
\usepackage{graphicx}
\usepackage{booktabs}
\usepackage{epstopdf}
\usepackage{gensymb}
\usepackage{aas_macros}
\usepackage{multirow}
\usepackage{lipsum}
\usepackage{url}
\usepackage[normalem]{ulem}
\usepackage[T1]{fontenc}
\usepackage[utf8]{inputenc}
\usepackage[english]{babel}
\usepackage[colorlinks=true,citecolor=blue,linkcolor=blue,urlcolor=blue]{hyperref}
\usepackage[capitalise,noabbrev]{cleveref}

\newcommand{\red}{\textcolor{black}}

\newcommand{\YT}[1]{\textcolor{black}{#1}}

\usepackage{tikz,xcolor}

\definecolor{lime}{HTML}{A6CE39}
\DeclareRobustCommand{\orcidicon}{\hspace{-4pt}
	\begin{tikzpicture}
		\draw[lime, fill=lime] (0,0) 
		circle [radius=0.16] 
		node[white] {\hspace{0.1mm}{\fontfamily{qag}\selectfont \tiny ID}};
		\draw[white, fill=white] (-0.07,0.1) 
		circle [radius=0.01];
	\end{tikzpicture}
	\hspace{-3.2mm}
}

\foreach \x in {A, ..., Z}{\expandafter\xdef\csname 
	orcid\x\endcsname{\noexpand\href{https://orcid.org/\csname orcidauthor\x\endcsname}
		{\noexpand\orcidicon}}
}

\begin{document}
	
    \title{{\color{black}Can Distance Duality Violation Save Late-time Solutions to the Hubble Tension?}}

    \author{Yashi Tiwari\orcidB{}}
    \email{tiwariyashi@itp.ac.cn}
    \affiliation{Institute of Theoretical Physics, Chinese Academy of Sciences (CAS), Beijing 100190, China}	
	
   \author{Ujjwal Upadhyay\orcidA{}}
    \email{ujjwalu@iisc.ac.in}
    \affiliation{Astronomy \& Astrophysics Group, Raman Research Institute,
    Bengaluru 560080, India}
    \affiliation{Department of Physics, Indian Institute of Science, 
    Bengaluru 560012, India}

    \author{Shao-Jiang Wang\orcidC{}}
    \email{schwang@itp.ac.cn}
    \affiliation{Institute of Theoretical Physics, Chinese Academy of Sciences (CAS), Beijing 100190, China}
    \affiliation{Asia Pacific Center for Theoretical Physics (APCTP), Pohang 37673, Korea}

    \author{Vivian Poulin\orcidD{}}
    \email{vivian.poulin@umontpellier.fr}
    \affiliation{Laboratoire Univers \& Particules de Montpellier,
    CNRS \& Université de Montpellier (UMR-5299), 34095 Montpellier, France}
    \footnotetext{The first two authors contributed equally to this work.\\}
		
	
\begin{abstract}
The discrepancy between early- and late-Universe determinations of the Hubble constant may point to physics beyond $\Lambda$CDM or to unaccounted-for systematics. Numerous late-time modifications to the expansion history have been proposed to alleviate this discrepancy, with limited success. Recent works have shown that, when the sound-horizon and supernova calibrations are held fixed, any purely late-time resolution requires a violation of the cosmic distance duality relation (CDDR). Recasting the tension in the $r_d$--$M_B$ plane, we show explicitly that distance duality, together with BAO and uncalibrated supernova data and a fixed sound-horizon calibration, determines $H_0$ independently of the late-time expansion history. We then test the viability of the required CDDR violation by separately constraining reciprocity violation and photon number non-conservation, deriving a new constraint on reciprocity-violating distortions of angular-diameter distances from BAO and cosmic-chronometer data. Combining this result with existing photon-number-conservation constraints, we find that the level of distance-duality violation needed to resolve the tension is strongly disfavoured by current data. We therefore conclude that, for fixed sound-horizon and supernova calibrations, no modification confined to the late-time expansion history---even one violating distance duality---can resolve the Hubble tension, pointing instead toward early-Universe physics or unresolved local systematics.

\end{abstract}

	
\maketitle
\section{ Introduction}
Despite the remarkable success of the $\Lambda$CDM model in describing a wide range of cosmological observations, increasingly precise \red{measurements have revealed significant tensions among parameters inferred from different cosmic epochs}. Most notably, the $>5\sigma$ discrepancy between the \red{local value} of the Hubble constant $H_0$, \red{inferred from} Cepheid-calibrated Type Ia supernovae (SNIa), and the value inferred from Planck CMB observations within \red{$\Lambda$CDM---the Hubble tension---} has emerged as a central challenge in contemporary cosmology. A broad spectrum of explanations, ranging from unaccounted-for systematic effects to extensions of the standard cosmological model involving new early- and/or late-time physics, has been proposed. Nevertheless, no compelling or broadly accepted resolution has emerged~(see~\cite{Cai:2026swf,Khalife:2023qbu,Hu:2023jqc, Abdalla:2022yfr, Schoneberg:2021qvd,Knox:2019rjx,DiValentino:2021izs,Verde:2019ivm} for recent reviews).

The Hubble tension can also be interpreted as a calibration tension, since $H_0$ is not measured directly but inferred from calibrated cosmological distance indicators such as SNIa (standard candles) and baryon acoustic oscillations (BAO; standard rulers), whose interpretation relies on external calibration parameters\red{~\cite{Aylor:2018drw}}. In the traditional distance ladder approach, Cepheid variable stars are used to calibrate the absolute magnitude $M_B$ of SNIa, which are subsequently used to infer the present-day expansion rate $H_0$~\cite{Riess:2021jrx}. In contrast, the BAO ruler scale, determined by the sound horizon $r_d$ imprinted in the CMB and large-scale structure, is calibrated from CMB observations within an assumed cosmological model, typically $\Lambda$CDM, thereby yielding a model-dependent inference of $H_0$ that is inconsistent with the distance ladder determination~\cite{Planck:2018vyg}. This discrepancy is equivalently a tension in the absolute magnitude $M_B$~\cite{Camarena2024, Efstathiou:2021ocp}: the value implied by the inverse distance ladder (CMB+BAO) conflicts with the Cepheid-calibrated one. Framed this way, the Hubble tension signals an inconsistency between the two distance probes, BAO and SNIa. In particular, once $r_d$ is fixed by the Planck 2018 $\Lambda$CDM best-fit, the tension manifests as an apparent violation of the cosmic distance duality relation (CDDR)~\cite{Poulin:2024ken, Teixeira:2025czm, Kanodia:2025jqh, Bansal:2026axl, Afroz:2025iwo}.

The CDDR establishes a direct relation between luminosity \red{and angular-diameter distances}, $d_L(z)=(1+z)^2d_A(z)$, which holds in any metric theory of gravity provided \red{photons propagate along null geodesics and their number is conserved}~\cite{etherington1933lx, 2007GReGr..39.1047E}. Consequently, any observed deviation would imply a breakdown of at least one of these assumptions and could provide evidence for physics beyond the standard cosmological framework~\cite{Bassett:2003vu, Uzan:2004my, Avgoustidis:2011aa, DeBernardis:2006ii, Lazkoz:2007cc, Ellis:2013cu, Holanda_2011, Lima:2011ye}. Motivated by this possibility, numerous studies have tested the CDDR using distance probes over broad redshift ranges and increasingly robust, calibration-independent techniques designed to minimize astrophysical and observational systematics~\cite{Holanda_2010, Li:2011exa, Khedekar:2011gf, Tang:2024zkc, Qi:2024acx, Yang:2024icv,Lopez-Hernandez:2025lbj}. Nearly all such analyses report no statistically significant evidence for a violation. This non-detection raises an important question: \red{if a CDDR violation is invoked to resolve the Hubble tension, why has the required signal not emerged in direct tests?} Recent studies have emphasized that, once the \red{SNIa absolute-magnitude calibration is held fixed}, any purely late-Universe resolution of the Hubble tension necessitates a violation of the CDDR~\cite{Teixeira:2025czm, Bansal:2026axl, Martinelli:2026wjp, Zhou:2025dxo}.

In this work, we revisit the $H_0$ tension through the lens of consistency between the two distance probes, guided by the CDDR. Treating distance duality as the fundamental anchor, we present a model-independent analysis that provides a transparent and intuitive reinterpretation of the tension and the viability of proposed resolution mechanisms. Focusing specifically on late-time solutions, we first demonstrate that, \red{for fixed calibrations}, imposing the CDDR effectively fixes the Hubble constant independently of the late-time expansion history, \red{so that reconciling local and inverse-distance-ladder determinations of $H_0$ requires} a substantial violation of distance duality. Further, we investigate whether such violations can be realized observationally by separately examining the two fundamental ingredients of CDDR: Etherington reciprocity and photon number conservation. We show that both possibilities are tightly constrained by existing data. In particular, we develop a complementary geometric probe of reciprocity-violating scenarios that distort the angular-distance sector, using transverse and line-of-sight BAO observations, independently of luminosity-distance measurements and conventional distance-duality tests. Taken together, these results indicate that the level of distance-duality violation required to alleviate the Hubble tension is strongly restricted by current observations, leaving little room for purely late-time solutions provided \red{$r_d$ and} the supernova $M_B$ calibration are fixed.

\noindent
\section{Connecting Hubble tension with CDDR}
\begin{figure}
    \centering
    \includegraphics[scale=0.6]{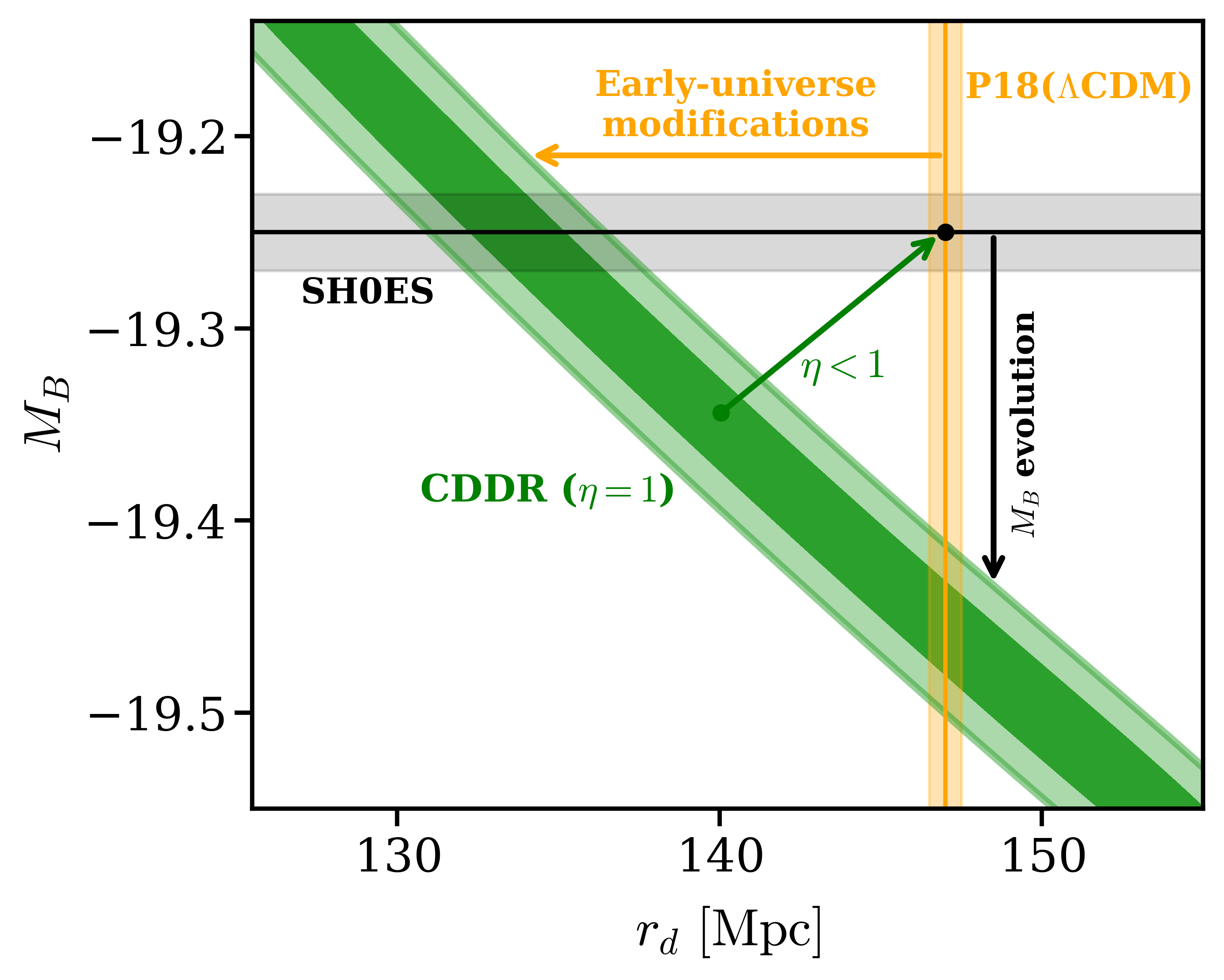}
    \caption{Constraint in the $r_d$--$M_B$ plane from the joint analysis of SNIa (Pantheon+) and BAO (DESI DR2) under solely assumption of the cosmic distance duality relation (CDDR). The green contour shows the allowed region from the combined BAO+SNIa data assuming $\eta=1$. The horizontal band denotes the $1\sigma$ SH0ES calibration of $M_B=-19.25\pm0.02$, while the vertical band represents the $1\sigma$ Planck 2018 $\Lambda$CDM determination of \red{$r_d=147.0\pm0.3\,{\rm Mpc}$}. The black point at \red{$(r_d, M_B)=(147\,{\rm Mpc}, -19.25)$} corresponds to the combination of the SH0ES and Planck calibrations ($H_0$ tension) and lies significantly away from the CDDR constraint. The arrows indicate the three principal directions in parameter space that can restore consistency: a reduction of the sound horizon $r_d$ (early-universe modifications), an evolution of the supernova absolute magnitude $M_B$ (local systematics), or a violation of the CDDR $(\eta<1)$.
    }
    \label{fig:rd-Mb constrain}
\end{figure}

The connection between the Hubble tension and the CDDR can be seen as an \red{early--late-Universe} calibration mismatch in the $r_d$--$M_B$ plane, as illustrated in Fig.~\ref{fig:rd-Mb constrain}. \red{A closely related calibration-based formulation in terms of the sound horizon was developed in Ref.~\cite{Aylor:2018drw}, while the $r_d$--$M_B$ perspective was discussed in detail in Ref.~\cite{Kanodia:2025jqh}.} We now derive this constraint and discuss its implications. The distance-duality parameter $\eta(z)$ is defined as
\begin{equation}
\eta(z) \equiv \frac{d_L(z)}{(1+z)^2 d_A(z)},
\label{eq: eta}
\end{equation}
where $\eta(z)=1$ corresponds to the validity of distance duality. The luminosity distance is inferred from supernova observations as
\begin{equation}
\red{\frac{d_L(z)}{\mathrm{Mpc}} = 10^{[m(z)-M_B-25]/5}},
\label{eq: D_L}
\end{equation}
where $m(z)$ is the observed apparent magnitude and $M_B$ sets the calibration. The angular diameter distance obtained from BAO measurements is given as
\begin{equation}
d_A(z) = \frac{r_d}{\theta_{\rm BAO}(z)(1+z)},
\label{eq: D_A}
\end{equation}
where $\theta_{\rm BAO}(z)$ is the observed angular scale of the BAO feature and $r_d$ sets the sound horizon calibration. Combining Eqs.~\eqref{eq: eta}–\eqref{eq: D_A} yields,
\YT{\begin{equation}
\eta(z,r_d,M_B)
=
\Bigg(\frac{10^{-(M_B+25)/5}}
{r_d/\mathrm{Mpc}}\Bigg)
\times
\Bigg(\frac{10^{m(z)/5\,}
\theta_{\rm BAO}(z)}{1+z}\Bigg).
\label{eq: eta-new}
\end{equation}}
\YT{The above equation makes explicit the calibration dependence of $\eta(z)$ when using SNIa and BAO observations. It shows that calibration
controls only the overall level of $\eta$, while its relative redshift dependence is fixed by the data
alone. Eq.~(\ref{eq: eta-new}) also indicates that while a constant CDDR violation can be degenerate with calibration choices, a redshift-dependent one cannot. Now, the above equation can then be rewritten as,}
\begin{equation}
\red{M_B + 5\log_{10}\!\left[\frac{r_d}{\mathrm{Mpc}}\,\eta(z)\right]}
= m(z) + 5 \red{\log_{10}}\!\left[\frac{\theta_{\rm BAO}(z)}{1+z}\right] - 25 .
\label{eq:constraint-equation}
\end{equation}
A key feature of Eq.~\eqref{eq:constraint-equation} is that its right-hand side depends on directly observed, uncalibrated SNIa and BAO data, while the left-hand side depends exclusively on the calibration parameters and possible deviations from distance duality. For simplicity, we assume a constant $\eta$, although $\eta$ may in principle be redshift dependent. Under this assumption, the above equation reduces to a constraint on the single combination,
\begin{equation}
\red{\mathcal{A} = M_B + 5\log_{10}\!\left[\frac{r_d}{\mathrm{Mpc}}\,\eta\right]},
\label{eq: CDDR-constraint}
\end{equation}
which encapsulates the fundamental degeneracy between the calibration parameters ($r_d$, $M_B$) and possible deviations from the CDDR ($\eta\neq1$). In the limit $\eta=1$, one recovers the standard CDDR-preserving combination $\mathcal{A}_0 = M_B + 5 \red{\log_{10}(r_d/\mathrm{Mpc})}$,
representing the consistency relation between the early-universe BAO calibration and the late-universe supernova calibration~\cite{Kanodia:2025jqh}. To demonstrate the $r_d$--$M_B$ degeneracy explicitly and connect with the Hubble tension, we perform an MCMC analysis, sampling the joint parameter space of $r_d$ and $M_B$ using the following likelihood function~\cite{Kanodia:2025jqh}:
\begin{equation}
\mathcal{L}_{\eta}(r_d,M_B)
\propto
\exp\!\left[
-\frac{1}{2}
(\boldsymbol{\eta}-\boldsymbol{\eta}_0)^{\mathrm T}
\mathbf{C}_{\eta}^{-1}
(\boldsymbol{\eta}-\boldsymbol{\eta}_0)
\right],
\label{e:likelihood}
\end{equation}
We fix $\boldsymbol{\eta}_0=\mathbf{1}$, corresponding to the validity of the CDDR. \YT{The vector $\boldsymbol{\eta}$ is defined as
$\boldsymbol{\eta}=\{\eta_i(r_d,M_B)\}$, where $i$ runs over the BAO redshift data points. It is inferred from the combined BAO and SNIa data using Eq.~(\ref{eq: eta-new}) and depends on the parameters $(r_d,M_B)$.}
Since BAO and SNIa measurements are not available at identical redshifts, we reconstruct the supernova luminosity distances at the BAO redshifts using Gaussian process regression (GPR), as described in Appendix~\ref{sec:app-A}. The covariance matrix $\mathbf{C}_{\eta}$ includes the full propagated uncertainties from both the GP reconstruction of the supernova luminosity distances and the BAO measurements (see Appendix~\ref{sec:app-A} for details).

In Fig.~\ref{fig:rd-Mb constrain}, we present the constraints in the $r_d$--$M_B$ plane obtained using the uncalibrated Pantheon+ SNIa sample~\cite{Scolnic:2021amr} and DESI DR2 observations~\cite{DESI:2025zgx}, under the assumption of the distance duality relation. 
The combination $\mathcal{A}_0 = M_B + 5\red{\log_{10}(r_d/\mathrm{Mpc})}$ is tightly constrained to $\mathcal{A}_0 = -8.61 \pm 0.02$. A key strength of this CDDR-based constraint is its minimal theoretical input: \textit{it is essentially independent of any cosmological model or parameterization, following directly from applying the CDDR to the uncalibrated BAO+SNIa data}. In contrast, the fiducial calibration underlying the Hubble tension---combining the $\Lambda$CDM-inferred sound horizon $r_d = 147 \pm 0.3$ Mpc with the SH0ES-calibrated absolute magnitude $M_B = -19.25 \pm 0.02$---yields $\mathcal{A}_{0,\rm fid} = -8.404 \pm 0.021$. This value disagrees with the CDDR-based constraint at the $7.0\sigma$ level. 
\par Figure~\ref{fig:rd-Mb constrain} offers key insights into the $H_0$ tension in a model-independent way. First, to remain consistent with the SH0ES calibration, the CDDR constraint forces $r_d$ to lower values relative to the Planck-$\Lambda$CDM fiducial value, suggesting a preference for early-time modifications. \red{Achieving the required reduction while preserving the successful CMB fit is, however, highly constraining and may require additional dynamics~\cite{Jedamzik:2020zmd}.} We do not pursue this direction further. We instead focus on late-time modifications proposed to resolve the $H_0$ tension. In such scenarios, pre-recombination physics remains unchanged, fixing $r_d$ to its CMB-inferred $\Lambda$CDM value. As shown in Fig.~\ref{fig:rd-Mb constrain}, achieving a higher $H_0$ (equivalently, shifting $M_B$ toward the SH0ES calibration) requires moving vertically at nearly fixed $r_d \simeq 147$ Mpc. This shift necessarily drives the solution away from the CDDR consistency band. Crucially, this conclusion is independent of the specific late-time expansion history invoked: \textit{\red{with fixed $r_d$ and standard distance duality, modifications confined to the late-time expansion history cannot resolve the Hubble tension~\cite{Aylor:2018drw}}}. Equation~\eqref{eq: CDDR-constraint} makes this result explicit: for fixed $r_d \simeq 147\,\mathrm{Mpc}$, $M_B$ (and hence $H_0$) is degenerate with $\eta$. Any shift in $M_B$ required to reconcile the tension must therefore be accompanied by $\eta \neq 1$, signalling a breakdown of the CDDR.
\par To make the connection between CDDR and the Hubble tension explicit, we note that the SH0ES distance ladder measurement effectively constrains the intercept of the low-redshift supernova Hubble diagram in a largely model-independent manner. This corresponds to constraining the combination
\begin{equation}
    \red{\mathcal{M}=M_B-5\log_{10}\!\left(\frac{H_0}{\mathrm{km\,s^{-1}\,Mpc^{-1}}}\right)},
    \label{eq: script_M}
\end{equation}
which encloses the degeneracy between the supernova absolute magnitude and the Hubble constant. Combining Eq.~\eqref{eq: CDDR-constraint} with Eq.~\eqref{eq: script_M}, we obtain
\begin{equation}
    \red{\frac{H_0}{\mathrm{km\,s^{-1}\,Mpc^{-1}}}
    =\frac{1}{\eta\,(r_d/\mathrm{Mpc})}\,
    10^{(\mathcal{A}-\mathcal{M})/5}}.
    \label{eq: inverse-ladder}
\end{equation}
\noindent
This is essentially the inverse-distance ladder~\cite{Camarena:2019rmj} generalised to allow for possible violations of CDDR. It shows that, for a given calibration of the sound horizon $r_d$, the combination of distance duality and the low-redshift supernova Hubble diagram determines $H_0$ independently of the late-time expansion history, \red{a point} also \red{emphasized} in Ref.~\cite{Efstathiou:2021ocp} in terms of the $M_B$ tension. Strictly speaking, Eq.~\eqref{eq: inverse-ladder} involves an effective redshift-averaged distance-duality parameter $\bar{\eta}$ when $\eta(z)$ is not constant. For notational simplicity, we denote this quantity by $\eta$ throughout. Assuming the validity of the CDDR, i.e. $\eta=1$, the uncertainty in the inferred value of $H_0$ arises solely from the observational uncertainties in $r_d$, $\mathcal{A}_0$, and $\mathcal{M}$:
\begin{equation}
\left( \frac{\sigma_{H_0}}{H_0} \right)^2
=
\left( \frac{\sigma_{r_d}}{r_d} \right)^2
+
\left( \frac{\ln 10}{5} \right)^2
\left( \sigma_{\mathcal{A}_0}^2 + \sigma_{\mathcal{M}}^2 \right).
\end{equation}
Adopting the measured value $\mathcal{M}=-28.5636\pm0.0088$~\cite{Riess:2016jrr,Efstathiou:2021ocp}, and the Planck-$\Lambda$CDM calibration of the sound horizon, $r_d=147.0\pm0.3~{\rm Mpc}$~\cite{Planck:2018vyg}, we obtain
\begin{equation}
    H_0= 66.6\pm0.7 \,\,\, \rm km\,s^{-1}\,Mpc^{-1},
\end{equation}
consistent with the standard inverse distance ladder determination. More generally, for fixed $r_d$, $\mathcal{A}$, and $\mathcal{M}$, any shift in the inferred value of $H_0$ must arise directly from a violation of distance duality satisfying,
\begin{equation}
\frac{\delta H_0}{H_0}
=
-\frac{\delta \eta}{\eta}\,.
\end{equation}
Therefore, reconciling the discrepancy between $H_0 \simeq 67$ and $H_0\simeq73~{\rm km\,s^{-1}\,Mpc^{-1}}$ would require $\delta H_0/H_0 \sim 0.1$, which translates into
$\delta \eta/\eta \sim -0.1$, i.e. an $\mathcal{O}(8\red{\text{--}}10\%)$ violation of \red{the} CDDR.\\

\noindent
\textbf{Possible ways forward:} 
The preceding analysis provides a useful framework to assess the viability of proposed resolutions to the Hubble tension. In particular, assuming the validity of the CDDR, late-time modifications to the expansion history cannot resolve the Hubble tension. We now summarize the remaining plausible directions and their limitations.

\begin{enumerate}
    \item Early-universe physics: The analysis of Fig.~\ref{fig:rd-Mb constrain} suggests that reducing the sound horizon (i.e., shifting the yellow band leftward) remains a viable route to alleviate the Hubble tension while remaining consistent with CDDR. Therefore, the resolution of the Hubble tension may ultimately involve at least some modifications to the early Universe~\cite{Vagnozzi:2023nrq}. However, our analysis relies primarily on background constraints; any viable early-time modification scenario must additionally satisfy constraints from CMB observations and other cosmological probes. 
    
    \item $M_B$ evolution: Another possible route to reconcile early- and late-universe measurements of $H_0$, while preserving CDDR, is to allow for an evolution in $M_B$ (corresponding to a downward shift of the grey band in Fig.~\ref{fig:rd-Mb constrain}). The most widely discussed possibility involves a sharp transition in $M_B$ at very low redshifts ($z\sim 0.01$), effectively decoupling the local SN calibration from the cosmological distance ladder~\cite{Bansal:2026axl}. However, such abrupt transitions are largely phenomenological and are often interpreted as indications of unresolved astrophysical or cosmological systematics. If one instead insists on a physical origin for the $M_B$ evolution, it can in principle arise from a late-time transition in $G_{\rm eff}$, as encountered in certain modified-gravity scenarios with non-minimal couplings~\cite{Marra:2021fvf,Ruchika:2023ugh, Ruchika:2024ymt}. An apparent $M_B$ evolution can equally arise from misinterpreting, within a standard $\Lambda$CDM analysis, a phantom transition in the dark energy at very low redshift. Nevertheless, both of these models are typically highly fine-tuned and severely constrained by multiple observational probes~\cite{Banik:2024yzi,Das:2026hfp,Hogas:2025mii}, except for the ultra-low-redshift $M_B$ (or equivalently $G_\mathrm{eff}$) evolution~\cite{Perivolaropoulos:2025gzo}, which can be regarded as an unaccounted-for local systematic.
    \item Violation of CDDR: A more radical possibility is a genuine breakdown of the CDDR itself, corresponding to an upward-rightward shift of the green band in Fig.~\ref{fig:rd-Mb constrain}. We discuss this scenario in detail in the next section, particularly exploring the observational implications of distance-duality violation.\\
\end{enumerate}

\noindent
\section{Viability of Distance Duality Violation?}
\label{sec: third}
The previous analysis leaves open the possibility that a violation of CDDR could provide a late-time resolution to the Hubble tension. In this section, we investigate the observational consequences of such violations and show that the level of deviation from distance duality required to resolve the $H_0$ tension is strongly constrained by existing observations.

\par
\begin{figure}
    \centering
    \includegraphics[scale=0.6]{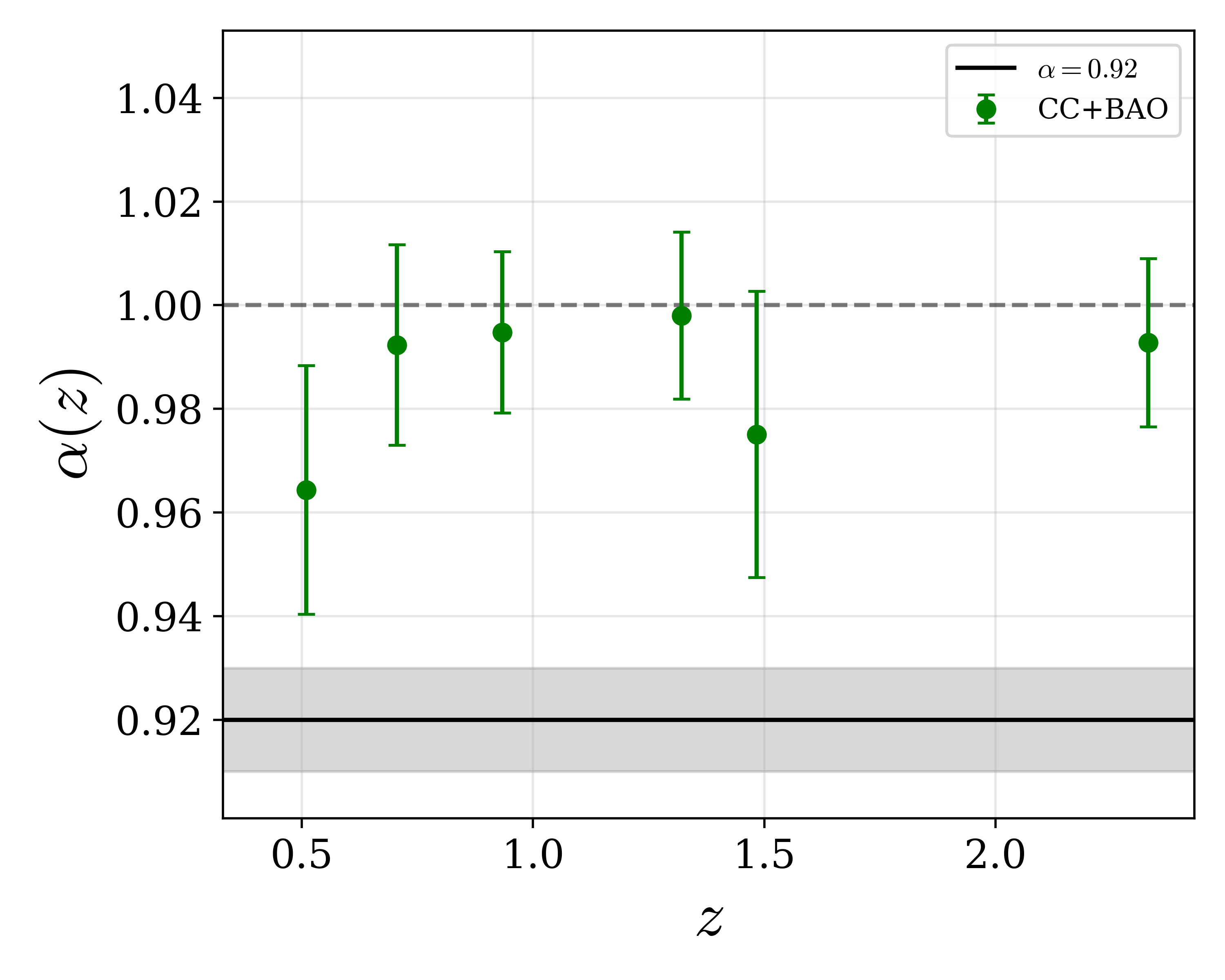}
    \caption{Constraints on reciprocity violations modifying the angular-diameter distance, derived from transverse and radial BAO measurements (DESI DR2) combined with cosmic-chronometer data~\cite{DESI:2025zgx, Moresco:2022phi}. \red{The grey band shows the value $\alpha=0.92$ required if the full distance-duality deviation is attributed to reciprocity violation ($\beta=1$); its width is the $1\sigma$ uncertainty propagated from the SH0ES measurement of $H_0$.}}
    \label{fig: alpha_constraint}
\end{figure}

 A breakdown of the CDDR can arise either from a violation of the Etherington reciprocity relation or from a breakdown of photon number conservation, or both. These two possibilities lead to qualitatively distinct observational signatures. Violations of reciprocity, as may occur in scenarios involving non-metric gravity or non-standard photon propagation, generically modify both the luminosity and angular diameter distances. In contrast, violations of photon number conservation affect only the luminosity distance while leaving the angular diameter distance unchanged. As discussed previously, and also noted in Ref.~\cite{Teixeira:2025czm}, any purely late-time resolution of the Hubble tension \red{under fixed calibrations} necessitates a significant departure from the CDDR, typically at the $\mathcal{O}(8\!-\!10\%)$ level. While the exact requirement depends on the assumed redshift evolution of $\eta(z)$, the required deviation is generically large. To assess the viability of this possibility, we separately investigate violations of the two assumptions underlying the CDDR---Etherington reciprocity and photon number conservation---and confront each with existing observational constraints.
\begin{itemize}
    \item \textbf{Etherington's reciprocity theorem:} Etherington's reciprocity theorem (ERT) relates the source angular distance $d_S$ and the observer area distance (angular-diameter distance) $d_A$ 
    in any metric theory of gravity provided photons propagate along null geodesics~\cite{2007GReGr..39.1047E}. Allowing for a violation of reciprocity, the relation can be written as
     \begin{equation}
        d_{A} = \frac{1}{\alpha(z)} \frac{d_S}{(1+z)} 
         \label{eq: reciprocity}
     \end{equation}
    where $\alpha(z)=1$ corresponds to the standard ERT. 
    The convention in Eq.~(\ref{eq: reciprocity}) is chosen such that, in the absence of photon number non-conservation, reciprocity violation alone is described by $\eta=\alpha$. In general, $\alpha\neq1$ can arise from a modification of the source distance or the angular-diameter distance. Since the source distance is not directly observable, reciprocity violations cannot, in general, be constrained independently of luminosity-distance measurements. However, a class of reciprocity-violating scenarios that modify the angular-diameter distance can be constrained geometrically. Considering $d_A=d_A^{\rm std }/\alpha(z)$, we can write
    \begin{equation}
        d_A(z)=\frac{\alpha(z)^{-1}}{1+z}\int_0^z\frac{dz'}{H(z')},
    \end{equation}
    where $\alpha(z)$ encodes a departure from the standard definition in a flat universe. Using transverse BAO measurements, one can write $d_A=\frac{r_d^{\rm th}}{1+z}\big(D_M/r_d\big)^{\perp}$\red{, where $r_d^{\rm th}$ denotes the adopted theoretical (Planck-$\Lambda$CDM) value of the sound horizon,} which yields
    \begin{equation}
    \label{eq: alpha_z}
    \alpha(z) = \frac{\int_0^z\frac{dz'}{H(z')}}{r_d^{\rm th}\left(\frac{D_{\rm M}(z)}{r_d}\right)^{\perp}}.
    \end{equation}
    Assuming that reciprocity violations affect only the transverse (angular) distance sector, while leaving the background expansion history and therefore the radial BAO distance $D_H=c/H$ unchanged, the degeneracy between $\alpha(z)$ and the expansion history can be removed. The reciprocity parameter can then be reconstructed directly from the transverse and radial BAO measurements,
    \begin{equation}
      \alpha (z)=\left[ \int_0^z \left(\frac{D_{\rm H}(z')}{r_d}\right)^{\parallel} dz'\right]\left[\left(\frac{D_{\rm M}(z)}{r_d}\right)^{\rm \perp}\right]^{-1},
      \label{eq: alpha-BAO}
    \end{equation}
    with $\perp$ and $\parallel$ denoting the transverse and radial BAO observables, respectively. 
    This relation constrains reciprocity-induced distortions of angular distances using BAO alone and is independent of the fiducial $r_d$ calibration. \YT{However, evaluating $\alpha(z)$ using Eq.~(\ref{eq: alpha-BAO}) requires the integral of $D_H/r_d$ from $z'=0$ to the redshift of interest. Since the available radial BAO measurements begin only at $z\sim0.5$, the low-redshift contribution to this integral, and consequently the overall normalization of $\alpha(z)$, is not sufficiently constrained by radial BAO alone. We therefore supplement the radial BAO measurements with cosmic chronometer (CC) data, which provide constraints on $H(z)$ at low redshifts and enable the reconstruction of the integrand over the full integration range. The role of the CC data in this regard is discussed in detail in Appendix~\ref{sec:app-A} and illustrated in Fig.~\ref{fig: alpha-detailed}. }Combining CC with BAO requires an $r_d$ calibration to convert between $H(z)$ and $D_H/r_d$; here we adopt $r_d=147\pm0.3$ Mpc. Thus, the BAO-only relation is calibration-independent, whereas the reported BAO+CC reconstruction is conditional on this calibration. \YT{In the future, low-redshift anisotropic BAO measurements, such as those from a BGS analysis, could provide $D_H/r_d$ directly in this redshift range, reducing the dependence on CC data and bringing the reconstruction closer to the calibration-independent, BAO-only form of Eq.~(\ref{eq: alpha-BAO}).} As shown in Fig.~\ref{fig: alpha_constraint}, BAO+CC observations~\cite{DESI:2025zgx, Moresco:2022phi} yield \YT{$\bar{\alpha}=0.990\pm0.008$}, with the tightest constraints around $z\sim1$, thereby strongly limiting reciprocity-violating modifications of angular distances. This construction is a complementary geometric probe that does not invoke luminosity-distance information or traditional CDDR tests.
    \par Furthermore, assuming photon number conservation, a nontrivial reciprocity parameter $\alpha(z)$ distorts the observed CMB spectrum from a perfect blackbody to a greybody, $I_{\nu,0}=\alpha(z_{\rm LSS})^{-1/2}I_{\rm BB}(\nu,T_0)$~\cite{Ellis:2013cu}. The COBE/FIRAS measurement of the CMB monopole spectrum therefore imposes the stringent constraint $|\alpha(z_{\rm LSS})-1|\lesssim5\times10^{-5}$ at recombination~\cite{Ellis:2013cu}, excluding any redshift-independent reciprocity violation. Together with the low-redshift bounds on $\alpha$ inferred in Fig.~\ref{fig: alpha_constraint}, these constraints strongly limit reciprocity-violating explanations of the Hubble tension.
    
    \item \textbf{Photon number conservation:} Assuming Etherington reciprocity remains valid, a violation of the CDDR can arise solely from photon number non-conservation during propagation. In this case, only the luminosity distance is modified:
     \begin{equation}
        d_{L} = (1+z) d_S \,\beta(z), 
        \label{eq: number-conversation}
    \end{equation}
    where $\beta(z)=1$ corresponds to photon number conservation. The convention in Eq.~(\ref{eq: number-conversation}) is chosen such that, in the absence of reciprocity violation (i.e. for $\alpha=1$), photon number non-conservation alone is described by $\eta=\beta$. Such processes generically alter the evolution of the CMB temperature, leading to a modified temperature--redshift relation, $T\propto(1+z)^{1-\epsilon}$, where $\epsilon$ captures the deviation from the standard scaling ($\epsilon=0$). This induces a redshift-dependent violation of distance duality which, under the standard assumption of adiabatic achromatic photon dimming, can be parameterized as~\cite{Avgoustidis:2011aa,Avgoustidis:2015xhk,Ruchika:2025sbb}
    \begin{equation}
    \beta(z)=(1+z)^{-3\epsilon/2}.
    \label{eq: beta}
    \end{equation}
Measurements of CMB temperature evolution through the thermal Sunyaev--Zel'dovich effect in galaxy clusters constrain $\epsilon$. We adopt the recent determination $\epsilon=-0.0106\pm0.0124$ from Ref.~\cite{Ruchika:2025sbb}. This translates into $\beta(1)=1.011\pm0.013$ and $\beta(2)=1.017\pm0.020$, showing that photon number non-conservation is constrained at the percent level over the redshift range relevant for late-time cosmology, well below the $\mathcal{O}(8\!-\!10\%)$ departure required to resolve the Hubble tension. \red{These bounds apply directly to microwave photons; extending them to the optical wavelengths relevant for SNIa requires the photon-number-violating mechanism to be approximately achromatic.}
\end{itemize}

\noindent
The above framework can be generalised to allow simultaneous violations of Etherington reciprocity and photon number conservation by combining Eq.~\eqref{eq: reciprocity} and Eq.~\eqref{eq: number-conversation}. In this case, the departure from distance duality is described by $\eta(z)=\alpha(z)\beta(z)$. Observational tests of the CDDR based on BAO, SNIa, \red{and cosmic chronometers} constrain this effective combination $\eta$, and are largely agnostic to the physical mechanism responsible for the violation. Recent model-independent tests of the CDDR based on Pantheon+/DESY5 supernovae and DESI BAO measurements find no statistically significant evidence for deviations from distance duality, typically constraining departures to the few-percent level over the redshift range relevant for late-time cosmology~\cite{Jesus:2024nrl, Yang:2025qdg}.

\par In Fig.~\ref{fig:cddr_constraints}, we summarise observational constraints on departures from the CDDR in the $\alpha$--$\beta$ plane, highlighting independent limits on reciprocity violations ($\alpha$) and photon number non-conservation ($\beta$), together with the corresponding joint constraint on $\eta=\alpha\beta$. The vertical allowed band (green regions excluded) shows the constraint on \YT{$\alpha(z_{\rm eff})$ derived from DESI BAO and CC observations around
$z_{\rm eff}=0.93$}, as presented in Fig.~\ref{fig: alpha_constraint}, \red{assuming the Planck-$\Lambda$CDM calibration of $r_d$}. The horizontal allowed band (grey regions excluded) shows the constraint on $\beta$ obtained from thermal Sunyaev--Zel'dovich measurements of $T_{\rm CMB}(z)$ in galaxy clusters~\cite{Ruchika:2025sbb}.
These analyses constrain the redshift-dependent deviation parameterised by Eq.~\eqref{eq: beta}, where we show the corresponding limits evaluated at $z\simeq1$. The diagonal band (purple regions excluded) denotes the observational constraint on $\eta$ obtained in Ref.~\cite{Jesus:2024nrl} from a cosmographic analysis combining SNIa, BAO, and cosmic chronometer data. We consider the \red{parameterization} $\eta(z)=1+\eta_0\frac{z}{1+z}$ and \red{translate the resulting constraint on $\eta_0$ into a bound on $\eta(z=1)$}. Although the precise width of this band depends on the assumed form of $\eta(z)$ and the dataset combination, the conclusions remain largely unchanged because the independent constraints on $\alpha$ and $\beta$ are substantially stronger. For comparison, we also show two representative levels of CDDR violation required to alleviate the Hubble tension: a constant offset, $\eta=0.92$, and a redshift-dependent profile, $\eta(z)=(1+z)^{-\frac{3}{2}\epsilon}$, with $\epsilon$ chosen to reproduce the required level of deviation at late times~\cite{Teixeira:2025czm}. All constraints are shown at the representative redshift $z\simeq1$, where both BAO and $T_{\rm CMB}(z)$ measurements provide meaningful constraints and where late-time modifications proposed to resolve the Hubble tension are expected to become observationally relevant. The figure shows that the magnitude of distance-duality violation required to reconcile early- and late-Universe measurements of $H_0$ is inconsistent with current constraints on $\alpha$ and $\beta$, effectively ruling out late-time CDDR violation as a viable solution to the Hubble tension.

\noindent
\section{Discussion}
The Hubble tension remains one of the most significant unresolved discrepancies in modern cosmology. Despite extensive theoretical and observational efforts over the past decade, no compelling consensus regarding its resolution has emerged. In particular, late-time modifications of the expansion history have repeatedly struggled to provide a satisfactory solution when confronted simultaneously with CMB, BAO, and SNIa observations \cite{DiValentino:2021izs, Zhou:2025kws, Pedrotti:2025ccw, Bansal:2026axl, Mukhopadhyay:2026fyk, Cai:2022dkh}. A common feature of such scenarios is that, although they often accommodate larger values of $H_0$ when constrained by CMB (and BAO) data, the inferred Hubble constant is driven back toward the Planck-$\Lambda$CDM value once supernova observations are included. In this work, we recast the $H_0$ tension in terms of possible violations of the CDDR and show that this behaviour is a direct consequence of the implicit assumption of CDDR, which acts as a powerful consistency condition linking luminosity and angular diameter distance measurements. In a model-independent framework, we demonstrated that the combination of BAO, SNIa, and the CDDR, \YT{along with the assumption of Planck-$\Lambda$CDM calibrated $r_d=147\pm0.3$ Mpc}, fixes the Hubble constant to $66.6\pm0.7\,{\rm km\,s^{-1}\,Mpc^{-1}}$, largely independent of the late-time expansion history and without relying on the Cepheid calibration of the supernova absolute magnitude. This explains why modifications that solely alter the late-time expansion history are unable to fully alleviate the Hubble tension. The implications of this result are broad. \red{Under the fixed-calibration and standard-propagation assumptions adopted here,} any scenario whose primary effect is to modify the recent expansion history, including dynamical dark energy, phantom-crossing models, interacting dark energy, and a wide class of late-time modified-gravity theories such as Horndeski models, will inevitably fail to alleviate the tension when confronted with combined CMB, BAO, and SNIa data.
\begin{figure}
\vspace{-0.5cm}
    \centering
    \includegraphics[scale=0.58]{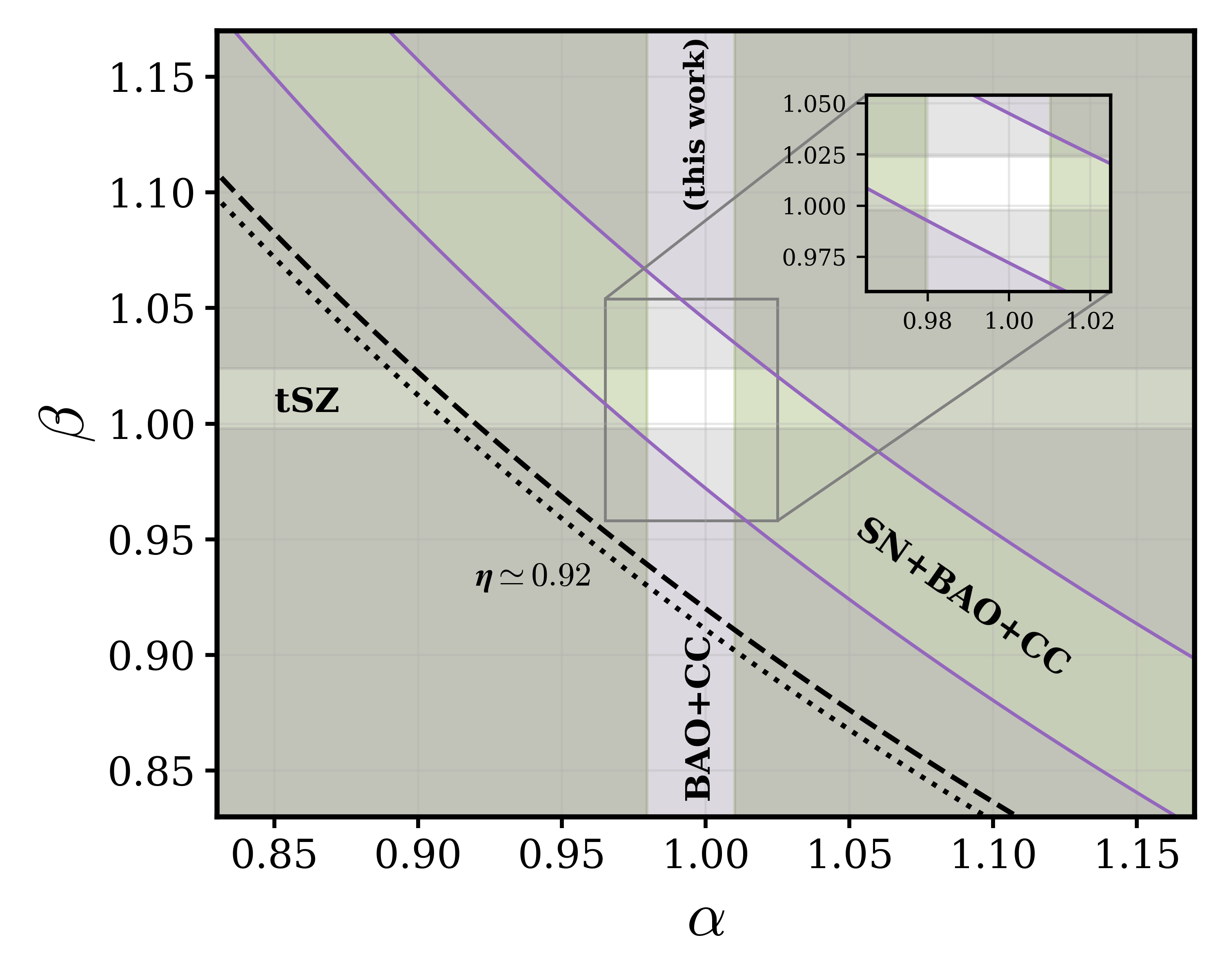}
    \caption{Constraints on departures from the cosmic distance duality relation (CDDR), parameterised as $\eta=\alpha\beta$, where $\alpha$ quantifies violations of Etherington reciprocity and $\beta$ encodes photon number non-conservation. The allowed horizontal, vertical, and diagonal bands show independent constraints on $\beta$, $\alpha$, and $\eta$, respectively, from different observations~\cite{Ruchika:2025sbb, Jesus:2024nrl, DESI:2025zgx, Moresco:2022phi}. The dashed and dotted lines indicate representative levels of CDDR violation required to alleviate the Hubble tension, corresponding to a constant $\eta\simeq0.92$ and a redshift-dependent profile $\eta(z)=(1+z)^{-\frac{3}{2}\epsilon}$. All constraints are shown at the representative redshift $z\simeq1$. The inset zooms into the intersection region, showing that the level of CDDR violation required to alleviate the Hubble tension is strongly disfavoured by current observations.
    }
    \label{fig:cddr_constraints}
\end{figure}
\par \red{In the second part of this work, we examined the two mechanisms through which distance duality can be violated---reciprocity violation and photon number non-conservation---and confronted each with present observations.} While constraints on photon-number non-conservation have been extensively discussed in the literature, direct low-redshift tests of reciprocity violation remain comparatively limited. We introduce a geometric probe of reciprocity-violating scenarios that distort angular-diameter distances while leaving the luminosity distance unaffected and show, using BAO and CC observations, that such departures are already tightly constrained by current data. Together, these constraints leave little room for the level of distance-duality violation needed to resolve the Hubble tension. 
\par It is worth noting that the constraints on photon-number non-conservation derived from the CMB temperature evolution rely on the assumption of adiabatic and approximately achromatic photon dimming. In principle, frequency-dependent mechanisms, such as photon--axion conversion, may evade such constraints. However, producing a strong departure from distance duality as required to alleviate the Hubble tension while remaining consistent with supernova colour measurements, quasar observations, opacity constraints, CMB data, and other tests of distance duality is highly challenging~\cite{Bassett:2003vu, M_nard_2010, Avgoustidis:2010ju, Avgoustidis:2015xhk, Holanda:2012ia, Ellis:2013cu}. On the other hand, reciprocity violations that affect only the source distance, and hence modify the luminosity distance without altering angular-diameter distances, evade the geometric constraint derived in this work. Nevertheless, they remain subject to conventional CDDR tests and are strongly constrained by current observations, as indicated by the diagonal band in Fig.~\ref{fig:cddr_constraints}. More broadly, this work provides a useful framework for assessing non-standard physics involving non-metric gravity, reciprocity violation, or photon number non-conservation, placing stringent constraints on their viability as solutions to the Hubble tension \cite{DAgostino:2026uct}. Understanding how specific theoretical realizations can satisfy the constraints presented here while remaining compatible with existing observations constitutes an interesting direction for future investigation.


\noindent
\section*{Acknowledgements}\label{sec:acknowledgments}\YT{We thank the anonymous referee for the constructive suggestions that helped improve the presentation of our manuscript.} YT and UU thank Shiv Sethi for insightful discussions and valuable comments during the early stages of this work. UU also thanks Suvodip Mukherjee for valuable feedback. SJW is supported by the National Key Research and Development Program of China Grants (No. 2021YFA0718304 and No. 2021YFC2203004), and the National Natural Science Foundation of China Grants (No. 12422502, No. 12547110, No. 12588101, No. 12235019, and No. 12447101). VP is supported by funding from the European Research Council (ERC) under the European Union's HORIZON-ERC-2022 (grant agreement no. 101076865). V.P.\ acknowledges the European Union's Horizon Europe research and innovation programme under the Marie Sk\l odowska-Curie Staff Exchange grant agreement no.\ 101086085 -- ASYMMETRY.\\

\noindent
\section*{Data Availability}The data that support the findings of this article are publicly available \cite{Scolnic:2021amr, DESI:2025zgx, Moresco:2022phi}.

\noindent
\appendix
\section{Methodology for Reconstruction of Luminosity Distance and $\alpha(z)$}\label{sec:app-A}
In this work, we employ Gaussian Process Regression (GPR) to reconstruct the luminosity distance $d_L(z)$ from supernova observations, required for the evaluation of $\eta_i$ in Eq.~\eqref{e:likelihood}, and to reconstruct the expansion history $H(z)$ from radial BAO and CC measurements, required for the evaluation of $\alpha(z)$ in Eq.~\eqref{eq: alpha_z}. GPR is a Bayesian non-parametric technique that reconstructs an underlying function directly from data by placing a prior over the space of functions rather than assuming a specific functional form~\cite{2020arXiv200910862W,10.5555/1162264,b60dec2e1b6c416387f33e9de784f573}. The reconstructed function is characterized by a covariance kernel and a set of associated hyperparameters, whose values are inferred from the data through maximization of the marginal likelihood. In this appendix, we summarize the \red{kernel and hyperparameter choices used in our analysis}.\\
 
{\bf General description of GPR:} Given a set of observations $\{z_i,f_i\}$, a Gaussian process describes the probability distribution function over the set of possible functions that can fit the observations, 
\begin{equation}
    f(z)\sim\mathcal{GP}(0,k(z,z')),
    \label{eq: GPR-def}
\end{equation}
where $k(z,z')$ is the covariance \red{kernel. We choose the Mat\'ern kernel with} smoothness parameter $\nu=3/2$~\cite{stein1999interpolation},
\begin{equation} 
k(z,z') = \sigma_f^2 \left(1+\frac{\sqrt{3}|z-z'|}{\ell}\right)\exp\left(-\frac{\sqrt{3}|z-z'|}{\ell}\right). 
\label{eq: GPR-kernel}
\end{equation}
Here, $\sigma_f$ and $\ell$ denote the amplitude and correlation length hyperparameters, respectively. We employ a custom implementation of GPR with a Mat\'ern kernel, where the kernel hyperparameters are optimized by maximizing the marginal likelihood using routines from the \texttt{scipy.optimize} package. Equations~\eqref{eq: GPR-def} and~\eqref{eq: GPR-kernel} essentially set the prior distribution over the function space to be updated to a posterior by the observations. The reconstructed function is described by a mean $\mu(z_*)$ and a variance $\sigma^2(z_*)$ given by
\begin{align} 
\mu(z_*) &= \mathbf{k}_*^\top (\mathbf{K} + \boldsymbol{\Sigma})^{-1}\mathbf{f},\\
\sigma^2(z_*) &= k(z_*,z_*) - \mathbf{k}_*^\top(\mathbf{K}+\boldsymbol{\Sigma})^{-1}\mathbf{k}_*, 
\end{align} 
where $\mathbf{K}_{ij}=k(z_i,z_j)$, $\mathbf{k}_*=[k(z_*,z_1),\ldots,k(z_*,z_n)]^\top$, and \red{$\boldsymbol{\Sigma}$ represents the observational covariance}.\\

\noindent
{\bf Reconstructing $d_L(z)$:} In order to apply GPR to reconstruct the luminosity distance $d_L$ from the Pantheon+ SNIa dataset~\cite{Scolnic:2021amr}, we need to convert the observed apparent magnitude to the luminosity distance, which requires the value of absolute magnitude $M_B$. Since $M_B$ is eventually treated as a free parameter in Eq.~\eqref{e:likelihood}, we reconstruct a rescaled luminosity distance $\bar{d}_L$, defined as 
\begin{equation}
\bar{d}_L(z)\equiv\frac{d_L(z)}{\lambda} = 10^{m(z)/5},
\end{equation} 
with a scaling factor $\lambda = 10^{-(M_B + 25)/5}$,
eliminating the dependence on $M_B$ from the reconstruction process. This enables a fast and flexible MCMC implementation for sampling in the $r_d$--$M_B$ space (see also Ref.~\cite{Kanodia:2025jqh} for more details). The full Pantheon+ covariance matrix is incorporated in the GPR reconstruction, yielding both the reconstructed mean luminosity distance and its covariance matrix. The reconstructed luminosity distance is then evaluated at the BAO redshifts to compute $\eta_i$ given by Eq.~(\ref{eq: eta}), while the reconstructed covariance is combined with the BAO covariance to construct the total covariance matrix $\mathbf{C}_\eta$,
\begin{equation}
\mathbf{C}_{\eta}
=
\mathbf{J}_{d_L}\,
\mathbf{C}_{d_L}\,
\mathbf{J}_{d_L}^{\mathrm T}
+
\mathbf{J}_{d_A}\,
\mathbf{C}_{d_A}\,
\mathbf{J}_{d_A}^{\mathrm T},
\label{eq:cov_eta}
\end{equation} where,
\begin{align}
(\mathbf{J}_{d_L})_{ij}
&=
\frac{\partial \eta_i}{\partial d_{L,j}}
=
\frac{\delta_{ij}}
{d_{A,i}(1+z_i)^2},
\\
(\mathbf{J}_{d_A})_{ij}
&=
\frac{\partial \eta_i}{\partial d_{A,j}}
=
-\frac{\eta_i}{d_{A,i}}
\,\delta_{ij}.
\end{align}
The covariance matrix $\mathbf{C}_\eta$ is then used in the likelihood of Eq.~\eqref{e:likelihood}. Sampling over the $(r_d,M_B)$ parameter space finally yields the constraints shown in Fig.~\ref{fig:rd-Mb constrain}.
\\
\begin{figure}
    \centering
    \includegraphics[scale=0.58]{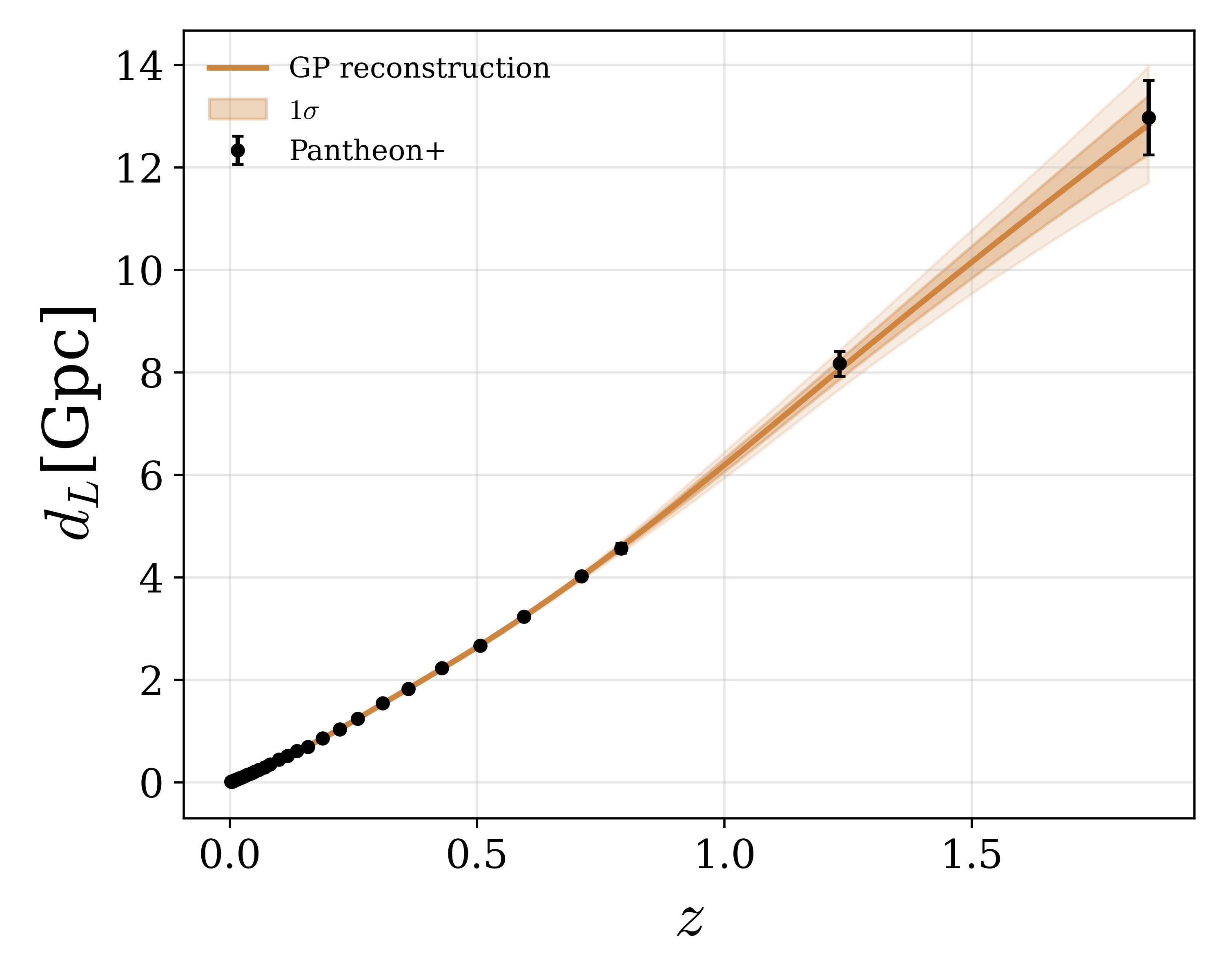}
    \caption{Gaussian Process reconstruction of the luminosity
distance $d_L(z)$ using the Pantheon+ SNIa for $M_B = -19.25$.
Black points represent binned data points, while the brown solid line indicates the GPR mean prediction. The shaded regions correspond to $1\sigma$ and $2\sigma$ uncertainty bands.
See Appendix A of Ref.~\cite{Kanodia:2025jqh} for further details.}
    \label{fig: d_L_rec}
\end{figure}
\begin{figure}
    \centering
    \includegraphics[width=1.0\linewidth]{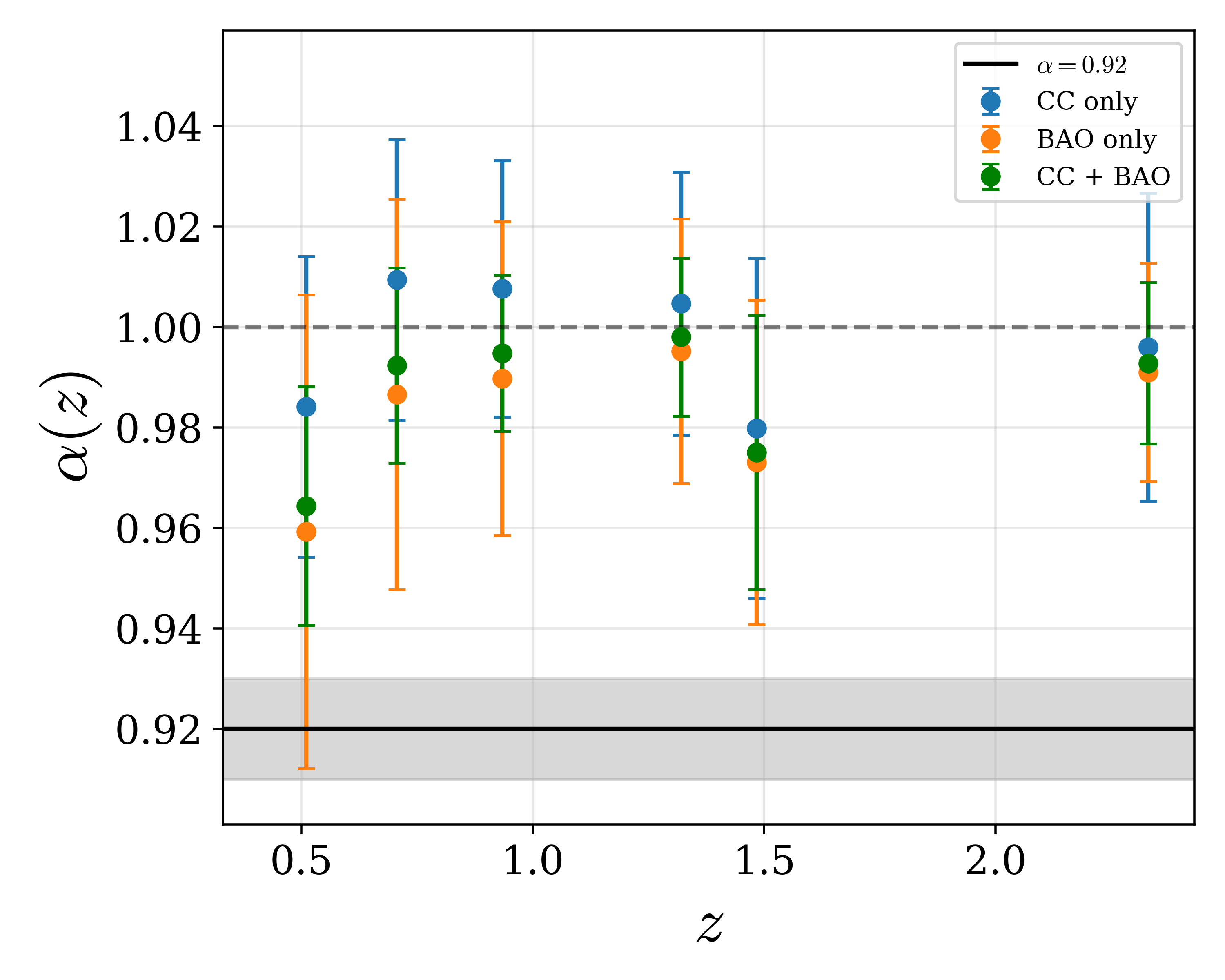}
    \caption{Reconstruction of $\alpha(z)$ from CC, BAO, and their combination. The BAO-only result marginalizes over the low-redshift $H_0$ anchor using $H_0\in[60,80]\,\,{\rm km\,s^{-1}\,Mpc^{-1}}$, while the CC-only and CC+BAO reconstruction adopts $r_d=147\pm0.3$ Mpc. The grey band shows the value $\alpha=0.92$ required if the full distance-duality deviation is attributed to reciprocity violation ($\beta=1$); its width is the $1\sigma$ uncertainty propagated from the SH0ES measurement of $H_0$.}
    \label{fig: alpha-detailed}
\end{figure}

\noindent
\YT{\bf Reconstructing $\alpha(z)$:} 
\YT{As discussed in Sec.~\ref{sec: third}, the reconstruction of the reciprocity parameter $\alpha(z)$, given by Eq.~(\ref{eq: alpha-BAO}), requires the integral of the reconstructed expansion history from $z'=0$ to the redshift of interest. Since the available $D_H/r_d$ measurements from radial BAO begin only at $z_{\rm eff}\simeq0.5$, the low-redshift contribution to this integral is not sufficiently constrained by BAO alone. Consequently, while the BAO measurements constrain the redshift dependence, or shape, of $\alpha(z)$, its overall normalization remains degenerate with the assumed low-redshift anchor $D_H(0)/r_d$, or equivalently with the choice of $H_0$ for a given $r_d$.}

\par \YT{To illustrate this dependence, we reconstruct $\alpha(z)$ for three cases: (i) using the BAO measurements alone, (ii) using the CC data alone, and (iii) using the combined CC+BAO data, as shown in Fig.~\ref{fig: alpha-detailed}. For the BAO-only reconstruction, the lack of low-redshift radial BAO measurements leaves the low-redshift anchor unconstrained. We therefore marginalize over $H_0\in[60,80]\,{\rm km\,s^{-1}\,Mpc^{-1}}$. As expected, this results in relatively large uncertainties in $\alpha(z)$, reflecting the sensitivity of the reconstruction to the low-redshift anchor. For the CC-only reconstruction, we reconstruct $H(z)$ and obtain $\alpha(z)$ using Eq.~(\ref{eq: alpha_z}), adopting the calibration $r_d=147\pm0.3$ Mpc. In this case, the low-redshift expansion history is directly constrained by the CC data, resulting in a better-constrained overall normalization and smaller uncertainties compared with the BAO-only case. The combined CC+BAO reconstruction gives constraints that lie between the CC-only and BAO-only cases, with the information from the CC data providing the low-redshift anchor while the BAO measurements further constrain the expansion history at higher redshifts. The comparison in Fig.~\ref{fig: alpha-detailed} therefore demonstrates that the low-redshift anchor affects both the overall normalization and the uncertainty of the reconstructed $\alpha(z)$. It also illustrates the role of the CC measurements in providing the low-redshift information required to constrain the normalization of $\alpha(z)$ in the absence of low-redshift radial BAO measurements.}

\par \YT{Nevertheless, our main conclusion remains unchanged: the reconstructed $\alpha(z)$ does not provide evidence for a significant violation of the distance duality relation of the magnitude required to address the Hubble tension. Although the lowest-redshift BAO-only point exhibits a relatively large uncertainty and can extend toward $\alpha\simeq0.92$, this isolated deviation should not be interpreted as evidence for reciprocity violation. In particular, the overall redshift dependence of $\alpha(z)$ does not show a coherent departure from $\alpha=1$ across the redshift range probed.}

\YT{The uncertainties in $\alpha(z)$ at different redshifts are correlated because the reconstructed comoving distance contains a common low-redshift contribution,
\begin{equation}
\chi(z)=\int_0^z \frac{c}{H(z')}\,dz'.
\end{equation}
Consequently, uncertainties in the low-redshift part of the reconstructed expansion history propagate coherently to $\alpha(z)$ at all higher redshifts, rather than affecting only the lowest-redshift points. In our reconstruction, these correlations are retained by using the same realization of the reconstructed $H(z)$ over the full redshift range when evaluating $\alpha(z)$ at different redshifts. The quoted uncertainties at individual redshifts therefore represent marginalized uncertainties and should not be interpreted as statistically independent. The corresponding CC+BAO reconstruction of $H(z)$, using GPR, is shown in Fig.~\ref{fig: H_rec}.
\begin{figure}[b!]
    \centering
    \includegraphics[scale=0.58]{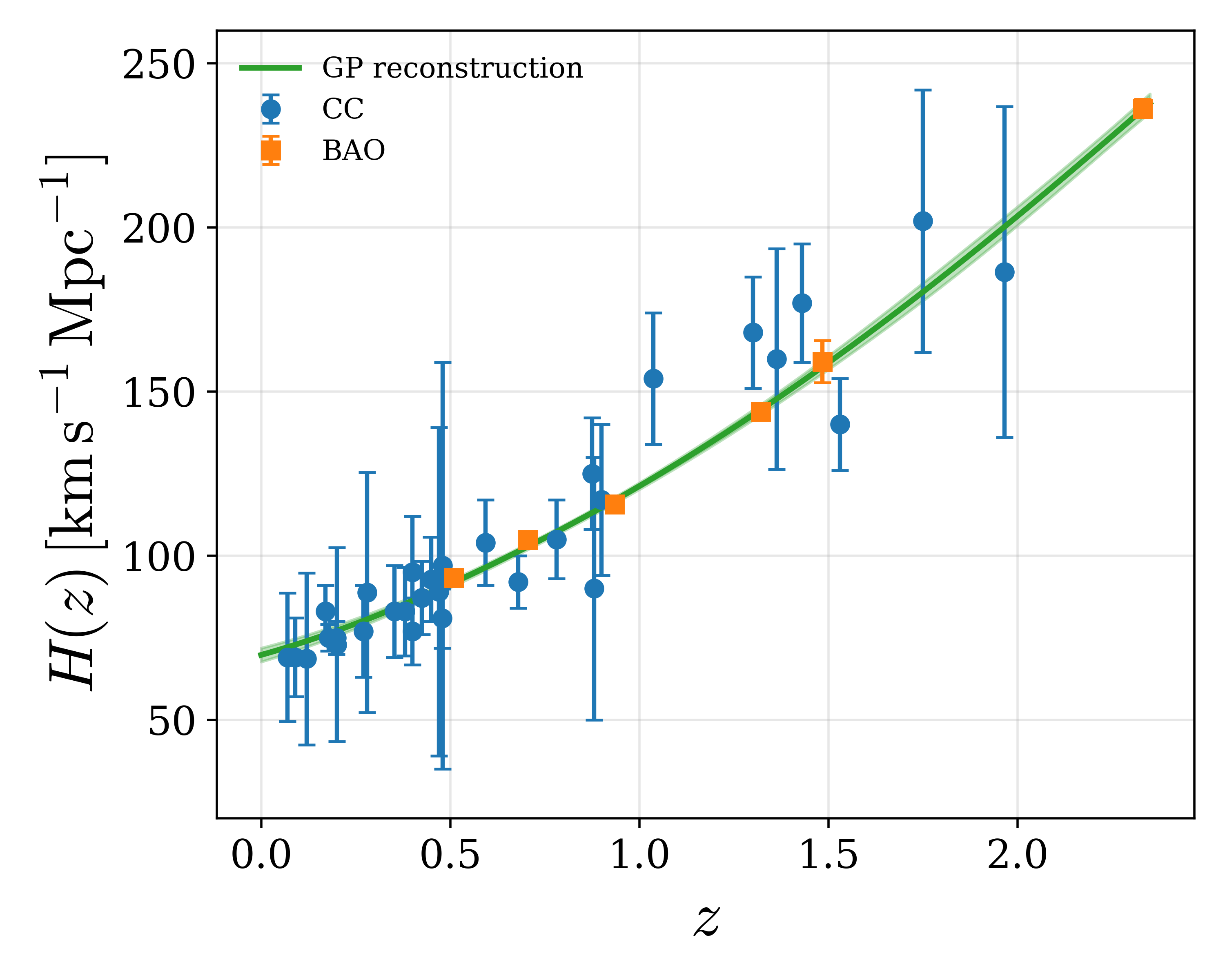}
    \caption{Gaussian-process reconstruction of $H(z)$ from radial BAO and CC measurements.}
    \label{fig: H_rec}
\end{figure}
The overall normalization of the reconstructed $\alpha(z)$ is also sensitive to systematic uncertainties in the CC measurements, including those associated with the choice of stellar population synthesis model used to estimate galaxy ages. Since the CC data provide important low-redshift constraints on the expansion history, such systematics can propagate coherently into the reconstructed $\alpha(z)$ and affect the inferred precision of the constraint. They should therefore be taken into account when interpreting the quoted uncertainty on $\alpha(z)$.}

\bibliography{ref.bib}

\end{document}